# The Effect of Dissipation on the Shapes of Dark Halos

JOHN DUBINSKI[1]

## ABSTRACT

The dissipative infall of gas during the formation of a galaxy modifies the density profile and shape of the dark halo. Gas dissipates energy radiatively and sinks to the center of the dark halo forming the luminous part of a galaxy. The resulting central density enhancement can alter the halo's orbital distribution. We simulate dissipative infall inside of an initially triaxial N-body dark halo by slowly growing a potential in the center of the particle distribution. We use both a disk and spherical potential to model the effect of spiral and elliptical galaxies respectively. The orbits of dark matter particles change in the presence of the central potential. Not only does the central density increase as predicted but the dark halo transforms from a prolate-triaxial halo ($T \sim 0.8$) to an oblate-triaxial halo ($T \sim 0.5$) while approximately preserving the flattening ($c/a \sim 0.5$). The main implication is that dark halos are rounder and more oblate than previous predictions of purely collisionless simulations with the new constraint that $b/a \gtrsim 0.7$. At the same time, the distribution of intrinsic flattenings of dark halos ($\langle c/a \rangle = 0.5$, $\langle (c/a)^2 \rangle^{1/2} = 0.15$) is preserved during the period of baryonic infall. The oval distortions of disk galaxies should therefore be slightly less than original predictions from collisionless dark halos. The predicted distribution of shapes for dark halos from cosmological N-body simulations is in better agreement with distribution of ellipticities of elliptical galaxies if we assume all halos and galaxies become oblate-triaxial in response to baryonic dissipation. The observed distribution of kinematic misalignment angles is also consistent with this shape distribution if elliptical galaxies rotate about their intrinsic minor axes.

## 1 INTRODUCTION

Galaxies are not spherical. The observed kinematics and apparent shapes of elliptical galaxies are consistent with a family of slowly rotating triaxial ellipsoids (Bertola & Cappaccioli 1975; Binney 1978; Franx et al. 1991). The spiral galaxies are well approximated by axisymmetric disks though many disks exhibit warps (e.g. Sanchez-Saavedra, Battener & Florido 1990) and there is evidence that disks are ovally distorted (e.g. Franx & de Zeeuw 1991) including the Galaxy with disk ellipticities as great as $\epsilon_{disk} = 0.1$ (Kuijken & Tremaine 1993). These effects naturally arise if the dark halo is triaxial. The polar ring galaxies also probe the flattening of the dark halo perpendicular to the plane of the disk through comparison of the rotation curve of the disk and the inclined ring (e.g. Whitmore, McElroy, & Schweizer 1987). The interpretation of the observations are model dependent but suggest a range of possible flattenings with $q \equiv c/a = 0.3 - 1.0$ (Sackett & Sparke 1991). The triaxiality of both elliptical galaxies and the dark halos of disk galaxies is a clue to the underlying physics of galaxy formation and these observations clearly place constraints on the different models of galaxy formation.

[1]Harvard–Smithsonian Center for Astrophysics, 60 Garden St., Cambridge, MA 02138
I: dubinski@cfa.harvard.edu

Triaxial dark halos occur naturally in hierarchical structure formation. The density peaks that are the site of galaxy collapse in the Gaussian random fields of hierarchical models are triaxial (Bardeen, Bond, Kaiser and Szalay 1986). Cosmological $N$-body simulations using the CDM power spectrum (Frenk et al. 1985; Dubinski & Carlberg 1991) and different initial power spectra (Warren et al. 1992) produce triaxial dark halos. These dark halos are highly flattened ($c/a \approx 0.5$) and often tend towards prolate-triaxial shapes ($c/b > b/a$), especially in their inner regions. The simulated dark halos also rotate slowly (dimensionless spin $\lambda \approx 0.05$) and the rotation occurs about an axis close to the intrinsic minor axis (Warren et al. 1992; Dubinski 1992). The shape distribution of simulated dark halos is not consistent with the distribution of apparent ellipticities (as measured from isophotes) for elliptical galaxies. The elliptical galaxies must come from an intrinsically rounder distribution though the distribution is not uniquely determined. While it is not obvious that the stellar component of elliptical galaxies and the presumed dark halo should have the same shape, it seems the most natural state of affairs if they form through the dissipationless merger of protogalactic substructure (Quinn, Salmon & Zurek 1988).

The missing ingredient in collisionless $N$-body simulations of dark halo formation is gaseous dissipation. In a complete picture of galaxy formation, baryons become segregated from the collisionless dark matter through dissipation of their gravitational energy through radiative cooling and star formation and sink to the center of the dark halo (e.g. White & Rees 1978). Different workers have begun to simulate galaxy formation using both gas and dissipationless dark matter quantifying this picture to some extent (Katz & Gunn 1991; Evrard, Summers & Davis 1992). The settling of a dense lump of baryons in the center can modify the dark halo. The central density of the dark halo increases in response to the added mass in the center changing the dark halo density profile (Blumenthal et al. 1986; Ryden & Gunn 1987; Ryden 1991; Flores et al. 1993). Collisionless N-body simulations do not include these effects. The density enhancement of the central baryons may also isotropize the orbits of the dark halo. A slowly rotating dark halo has a large fraction of box orbits that by their nature pass near the center. A central density enhancement can alter the distribution of orbits and therefore modify the shape of the dark halo. Recent models of galaxy formation that include gas (Evrard et al. 1993) produce a much rounder and more oblate distribution of dark halos than purely dissipationless models.

In this paper, we examine the effect of dissipation on the shapes of dark halos. We simulate dissipative infall by growing a central potential corresponding to a luminous galaxy in the center of a dark halo generated with a cosmological N-body simulation. The orbits of the dark halo adjust to the presence of the central potential and form a more oblate distribution while preserving the initial flattening. We therefore expect a larger fraction of oblate galaxies than predicted by dissipationless simulations. If the effect is universal, we can produce an intrinsic ellipticity distribution of dark halos that is consistent with the elliptical galaxies contrary to predictions from collisionless simulations. In §2 we describe the simulations. In §3 we present the results on the modified shapes and density profiles of dark halos. In §4 we discuss the importance of the change in shape to the internal properties of elliptical and spiral galaxies and provide an educated guess of the intrinsic distribution of elliptical galaxies. In §5 we summarize our findings.

## 2 The Experiments

### 2.1 The Dark Halo

The first requirement of our experiments is a representative dark halo. We set up the initial conditions for the dark halo using the technique described in Dubinski & Carlberg (1991) and generate the halo using an N-body simulation. Our simulation assumes a standard CDM power spectrum ($H_o = 50$ km s$^{-1}$, $h = 1/2$, $\Omega = 1.0$) normalized so that $\sigma_8 = 1.0$ at the present epoch where $\sigma_8$ is the variance in mass excess in spheres of radius $R = 8h^{-1}$ Mpc. In brief, we choose a density peak from a Gaussian random field of density fluctuations. We set up the random field inside of cube that is 8 Mpc (128 divisions) on a side by Fourier transforming a random sampling of waves from the CDM power spectrum. The density field is convolved with a Gaussian filter with a filter radius, $R_f = 0.75$ Mpc corresponding to a galactic scale. We choose a peak with a height in the range $\nu = 2.0$ to $3.0\sigma$ that is isolated from its neighbors. Although, density peaks are not necessarily collapse sites because of the effect of tidal shear (Katz et al. 1993, Bertschinger & Jain 1993), high peaks often are and the peak discussed here is a collapse site. We set up the initial conditions by laying particles in a cubical lattice within a sphere surrounding the peak and calculate the positional shifts and peculiar velocities using the Zel'dovich (1970) approximation . We also add a Hubble velocity, $\mathbf{v} = H\mathbf{r}$, to each particle. We account for the external torque with a time-dependent tidal potential corresponding to a $1\sigma$ tidal field as described in Dubinski & Carlberg (1991).

We simulate the halo using an $N$-body tree code based on the Barnes-Hut algorithm (Barnes & Hut 1986) employing a multiple timestep scheme similar to the method described by Hernquist and Katz (1989). Each particle is assigned an individual timestep from a hierarchy of values that differ by factors of 2 according to the estimate of the local time scale. The particle timestep is estimated from the value of its current acceleration through $\Delta t = Ca^{-\alpha}$ where $a$ is the acceleration and $C$ and $\alpha$ are arbitrary constants. We use $\alpha = 0.75$ which is a compromise between the expected time scale dependence for an isothermal sphere ($\alpha = 1/2$) and a point mass ($\alpha = 3/2$). We select a constant, $C$, which is small enough to minimize integration error but as large as possible to save computing time. As the acceleration on the particle increases or decreases the timestep is either multiplied or divided by a factor of two. For the dark halo simulations, the minimum time step is set to $\Delta t = 2.8$ Myr and ranges in powers of 2 up to a maximum value of $\Delta t = 90$ Myr with the constant $C = 22.5$ Myr.

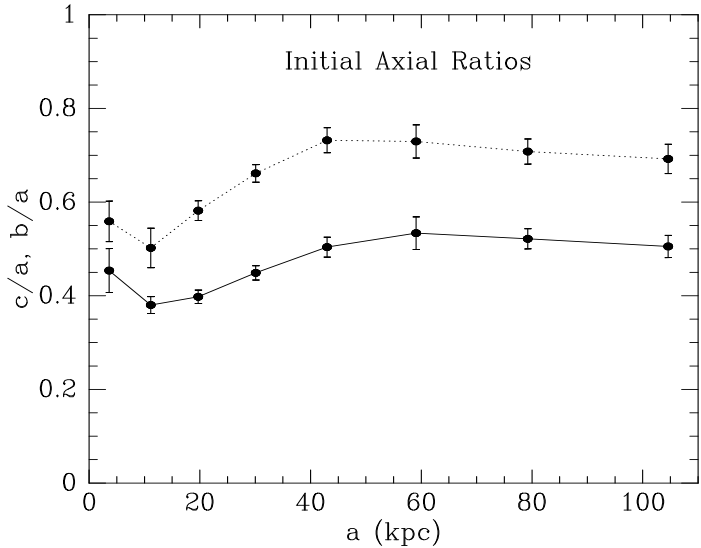

FIG. 1.—Axial ratio profiles of the dark halo. The solid and dotted lines respectively trace the ratios $c/a$ and $b/a$. The error bars represent the variance in the axial ratios for measurements from 20 time intervals spanning 2.0 Gyr at the end of the simulation. The errors in the axial ratios are small showing that the dark halo is in equilibrium.

The N-body simulation uses approximately 70000 particles arranged in a sphere with a diameter of 2 comoving Mpc containing a total mass of $2 \times 10^{12}$ M$_\odot$. The particle softening radius is 1.1 kpc. The initial peak height is $\nu = 2.5\sigma$. We ran the simulation until $\sigma_8 = 0.5$ until the halo was virialized. The final dark halo is a highly flattened nearly prolate object as shown by its initial axial ratio profile (Figure 1). We calculate the axial ratio profile using the method described in Dubinski & Carlberg (1991) that relates the shape of ellipsoidal shells to the dimensionless inertia tensor, $I_{ij} = \sum x_i x_j / a^2$ where $a^2 = x^2 + y^2/q_1^2 + z^2/q_2^2$ is the elliptical radius. We determine the axial ratio in independent ellipsoidal shells containing roughly equal amounts of mass from $r = 5$ to $100$ kpc. There are approximately 5000 particles in each bin so the errors are small ($< 5\%$ in the axial ratio [Dubinski & Carlberg 1991]). We demonstrated the stability of the halo by running each simulation for a further 2 Gyr and calculated the axial ratio profiles at 20 intermediate steps. The error bars in Figure 1 show the variance in the axial ratios at each radius by averaging over 10 consecutive snapshots.

The density profile of the dark halo is well described by an ellipsoidal Hernquist potential of total mass $M = 2.8 \times 10^{12}$ M$_\odot$ and elliptical radius $a = 28$ kpc. The maximum circular velocity for the dark halo is $v_{max} = 320$ km s$^{-1}$. We decided to rescale the mass of the halo so that $v_{max} = 200$ km s$^{-1}$ for the models below while leaving the scale of $a$ unchanged.

The total mass of the rescaled halo is therefore $M_{halo} = 1.1 \times 10^{12}$ M$_\odot$. The model therefore corresponds to a bright galaxy like our own. The dimensionless spin parameter is $\lambda = 0.01$ for this model a small value but still representative. The angular momentum vector aligns with the minor axis of the potential within 10 degrees.

## 2.2 Galactic Potentials

Galaxy formation in the context of hierarchical structure formation is a messy problem. In hierarchical models, a protogalaxy is composed of a collection of smaller scale fluctuations (the substructure), that eventually merge to form a smooth, centrally concentrated galaxy. The process of dissipative infall has been traditionally discussed in reference to a single monolithic collapse of a spherical perturbation (e.g. Rees & Ostriker 1977; White & Rees 1979). Gravitational collapse and dissipation occur concurrently and if the dissipative cooling time $t_{cool}$ is less than the gravitational freefall time, $t_{ff}$ the gas will sink to the center, forming the luminous galaxy. However, in a hierarchical collapse gas first dissipates within the lumps of substructure and is perhaps later reheated when merging occurs. During the time of merging, the gas is therefore probably cycled through hot and cold phases many times. Only after the massive mergers of the substructure are complete in the collapse, can the gas finally cool undisturbed and settle to form the galaxy. The mass cooling rate, $\dot{M}$ for bright galaxies ($V_c \sim 250$km s$^{-1}$) is $\sim 100$M$_\odot$ yr$^{-1}$ (White & Frenk 1991) so the cooling time for a baryonic mass of $10^{11}$ M$_\odot$ is around 1 Gyr.

We therefore model the formation of the galaxy by slowly growing a potential in the center of a relaxed dark halo over a period $\sim 1$ Gyr. (e.g. Barnes & White 1984). We grow the mass of the potential at a linear rate until it reaches a maximum mass comparable to the expected luminous mass of galaxies. We have chosen two growth times of $t_{growth} = 1.6$ Gyr and 3.2 Gyr corresponding to 8 and 16 rotation times at a $R = 10$ kpc. The galaxy potential is treated as an individual particle within the simulation with its unique force law determined by the functional form of the potential. The potential can therefore move around within the simulation. This avoids the problem of the dark halo drifting away from the center of a rigid potential. In practice, the freely moving potential quickly settles to the dense center of the dark halo because of dynamical friction.

We use two forms of the potential to model both disk and spherical galaxies to represent the effects of spiral and elliptical galaxies. For the disk potential we use the Kuzmin-Toomre (KT) disk (Binney & Tremaine 1987),

$$\Phi = -\frac{GM_{KT}}{(R^2 + (R_{KT} + |z|)^2)^{1/2}} \tag{1}$$

mainly because of its simplicity. (In practice, we replace $|z|$ with $(z^2 + b^2)^{1/2}$ where $b$ is the softening radius in the N-body simulation so the potential is actually that of Miyamoto & Nagai (1975)). The rotation curve of KT disk resembles the rotation curve of an exponential disk when $M_{KT} = 1.3M_e$ and $R_{KT} = 1.3R_e$ where $M_e$ is the mass of an exponential disk and $R_e$ is the scale length. We wish to set up a model with a combined disk and halo potential that resembles the rotation curve of a real galaxy. We choose a ratio of dark mass to baryonic mass ratio equal to one within 3 exponential scale radii, a situation similar to our own galaxy (e.g. Kuijken 1989). The exponential scale radius for the disk is $R_e = 3.5$ kpc (the value for the Galaxy) so our equivalent KT disk potential has a scaling radius, $R_{KT} = 1.3 R_e = 4.6$ kpc. We then equate the rotation curve velocity at 3 scaling radii to the maximum value for the dark halo, $v_{max} = 200$ km s$^{-1}$. The total mass of the KT disk is therefore $M_{KT} = 1.1 \times 10^{11} M_\odot$ The composite halo galaxy model has a maximum rotation curve velocity around 280 km s$^{-1}$ and is a reasonable representation of real rotation curve. We align the potential so that the disk axis aligns with the minor axis of the dark halo. We might expect this alignment since the sense of rotation of the disk is probably similar to the dark halo. However, SPH simulations of galaxy formation show that large misalignments between the disk and dark halo minor axis are possible (e.g. Katz & Gunn 1991).

For the elliptical galactic potential we use the spherical Hernquist (1990) potential,

$$\Phi = -\frac{GM_H}{r + r_H}, \qquad (2)$$

which is a good fit to the deprojected de Vaucouleur $R^{1/4}$ law of elliptical galaxies. The typical effective radius for bright ellipticals is around $R_e = 6$ kpc ($h = 1/2$) (Binney and Tremaine 1987), corresponding to a Hernquist radius of $r_H = 3.2$ kpc [since $R_e = 1.82 r_H$ from Hernquist (1990)]. We use a spherical potential to model what is intrinsically an ellipsoidal distribution mainly because of simplicity but also as a minimal hypothesis since we do not know the initial relative orientation of the stellar component and the dark halo. Presumably in a self-consistent system they would relax to the same shape during the period of infall. Unlike spiral galaxies, the ratio of luminous to dark matter is generally unknown in ellipticals so we are forced to make a reasonable guess (although see Saglia et al. 1993). We set the peak value of the "rotation curve" of the elliptical galaxy that occurs at $r = r_H$ to $v = 200$ km s$^{-1}$, the peak value for the dark halo. Therefore, the mass of the galaxy potential is about the same as the mass of the dark halo within a few effective radii similar to some recent observational estimates based on the velocity dispersion profiles in elliptical galaxies (e.g. Saglia et al. 1993). The total mass of the luminous galaxy for the composite model is $M_H = 1.4 \times 10^{11}$ M$_\odot$.

## 3   RESULTS

We did four simulations: two with a Hernquist sphere growing over times of 1.6 and 3.2 Gyr, and two with a KT disk growing over the same times. After the galactic potential finished growing, we continued the simulations for a further 1.6 Gyr. We find that there is no significant difference between the final results for the two different growth times so we only discuss those models with $t_{growth} = 1.6$ Gyr.

Figure 2 and 3 show the spherically averaged density profiles and rotation (circular velocity) curves for the composite models at the end of the simulations. We note that the the final dark halo has been compressed by the galactic potential (Fig. 2) as discussed by several authors (e.g. Blumenthal et al. 1986; Flores et al. 1993; Ryden 1991). The central density increases by a factor of 2 for the disk model and an a factor of 3 for the spherical model. The resulting rotation curves are reasonably flat and plausible models for real rotation curves of galaxies although perhaps one might argue that they are falling too rapidly (Fig. 3).

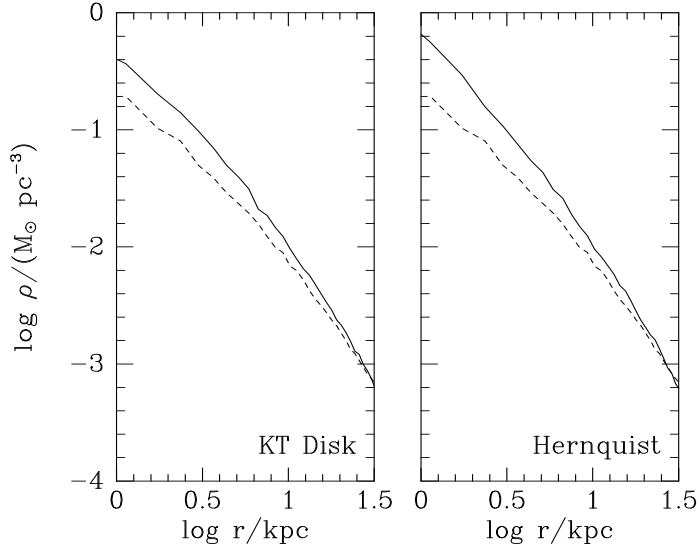

Fig. 2.–Spherically averaged density profiles of the final composite models. The profiles of the dark halo and the combined galaxy/dark halo model are presented by the dashed and solid lines respectively.

Figure 4 presents the change in the axial ratio profiles and triaxiality parameter, $T = (a^2 - b^2)/(a^2 - c^2)$ (Franx et al. 1991) in response to the growing central potential. The profiles are the average of values found at 10 time intervals spanning the additional 1.6 Gyr of simulation time after the end of the growth of the galactic potential. The error bars show the variance of the measured values in these 10 intervals. The small size of the error bars indicates that the composite halo models are stable after the central potential has stopped growing. In the disk potential model, the flattening $c/a$ remains virtually unchanged (although there is a slight increase) whereas the $b/a$ axial ratio increases from a value of 0.5 to 0.7 in the inner regions and to 0.8 at larger radii. In the Hernquist potential model, the effect of the potential is stronger. The ratio $c/a$ increases from $c/a = 0.4$ to $c/a = 0.6$ while $b/a$ increases to 0.8 over the entire halo. The halo triaxiality changes from about $T = 0.9$ (prolate) initially to $T = 0.6$ for the disk model and $T = 0.5$ for the spherical model. Figure 5 also shows the transformation to an oblate shape by presenting the "isophotes" in the plane of intermediate and major axis (view down the minor axis) before and after the growth of central potential.

The growth of the central potential makes the dark halo more oblate (although the final value of $T \approx 0.5$ corresponds to a "maximally" triaxial ellipsoid) and rounder for the spherical potential. We can understand this effect by considering influence of a central potential on the orbits of particles within the dark halo. The analysis of the effect of baryonic infall on dark halos has usually been done using spherical potentials (e.g. Blumenthal et al. 1986; Ryden 1991; Flores et al. 1993). Under the assumption of a slowly growing galaxy, there are adiabatic invariants of the dark halo orbits namely the angular momentum and the radial action. For slowly rotating triaxial potentials, the orbits are mainly box orbits though the presence of rotation about the minor axis and the roughly ellipsoidal shape indicates that there are also some tube orbits. The box orbits at all radii will pass near the center of the dark halo and should therefore interact with the central galactic potential. Figure 4 reveals that the halo axial ratio profile changes all the way out to $a \sim 100$ kpc. The changes in shape at all radii show that the galactic potential is strongly affecting the shapes of the box orbits.

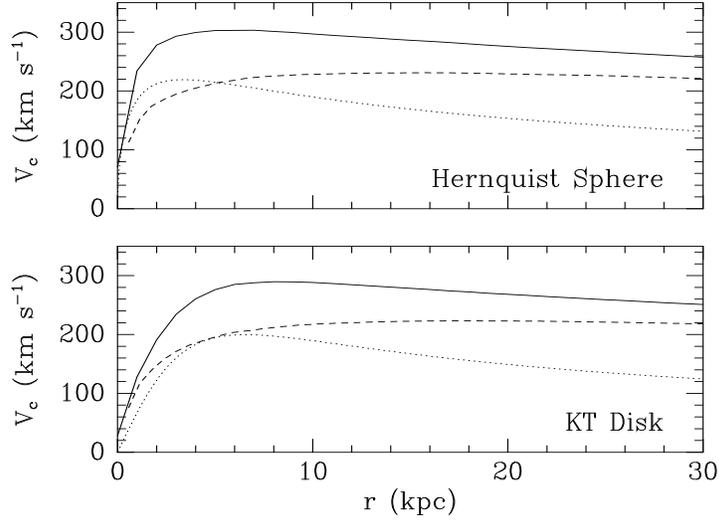

FIG. 3.–Rotation curves of the final composite models. The rotation curves of the galactic potential, the dark halo and the combined galaxy/dark halo model are presented by the dotted, dashed and solid lines respectively.

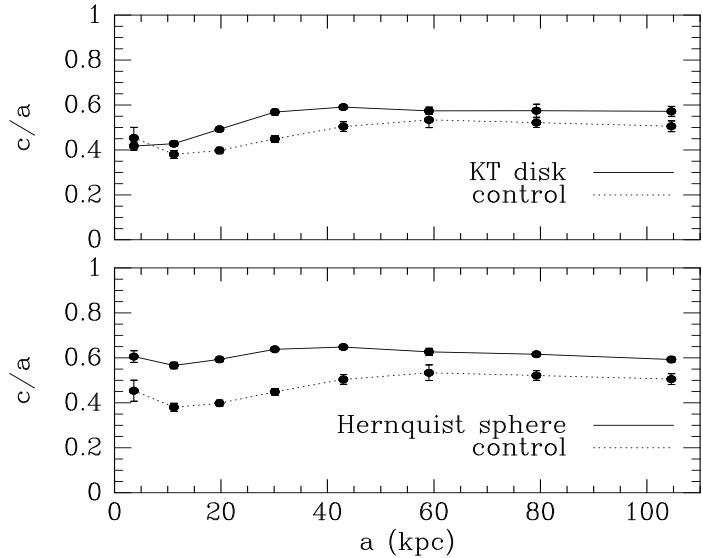

FIG. 4a.–Evolution of the (a) $c/a$ ratio (b) $b/a$ ratio and (c) triaxiality $T$ for the growing KT disk and Hernquist sphere. The dashed line presents the results for the control experiment in which no central potential was included. The solid line present the results including the potentials. The error bars show the variance in shape measurements for 10 time intervals after the central potential has stopped growing. The small error bars show that the models are in equilibrium.

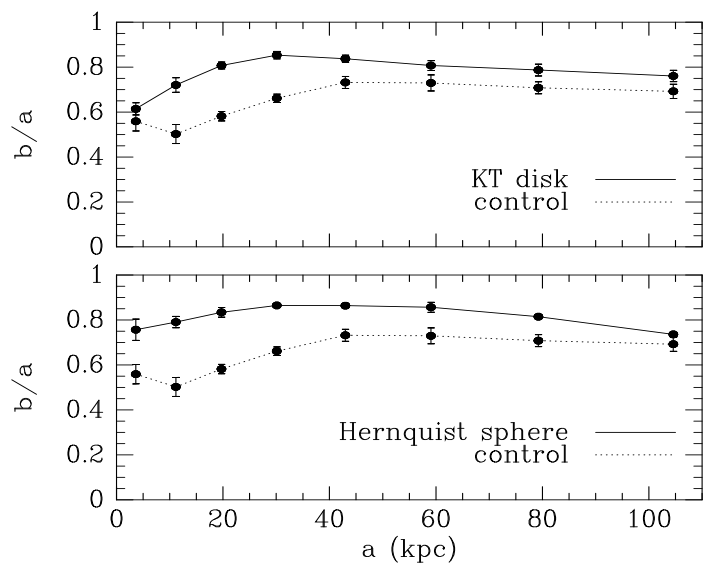

FIG. 4b.–

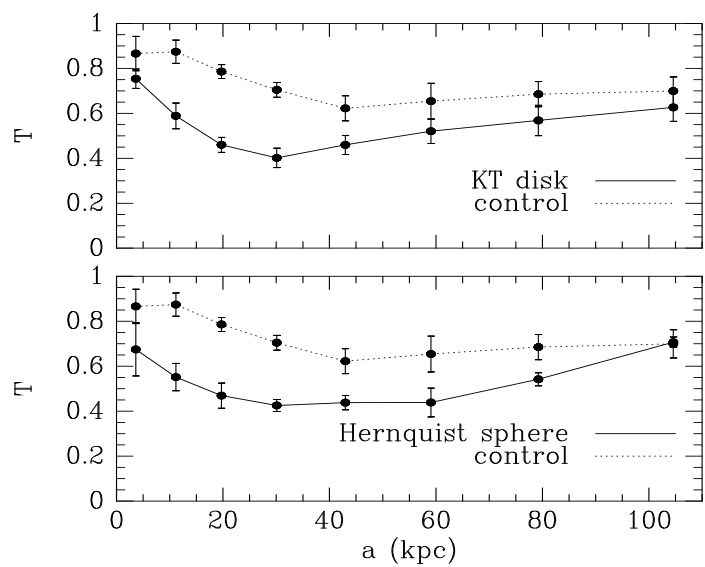

FIG. 4c.–

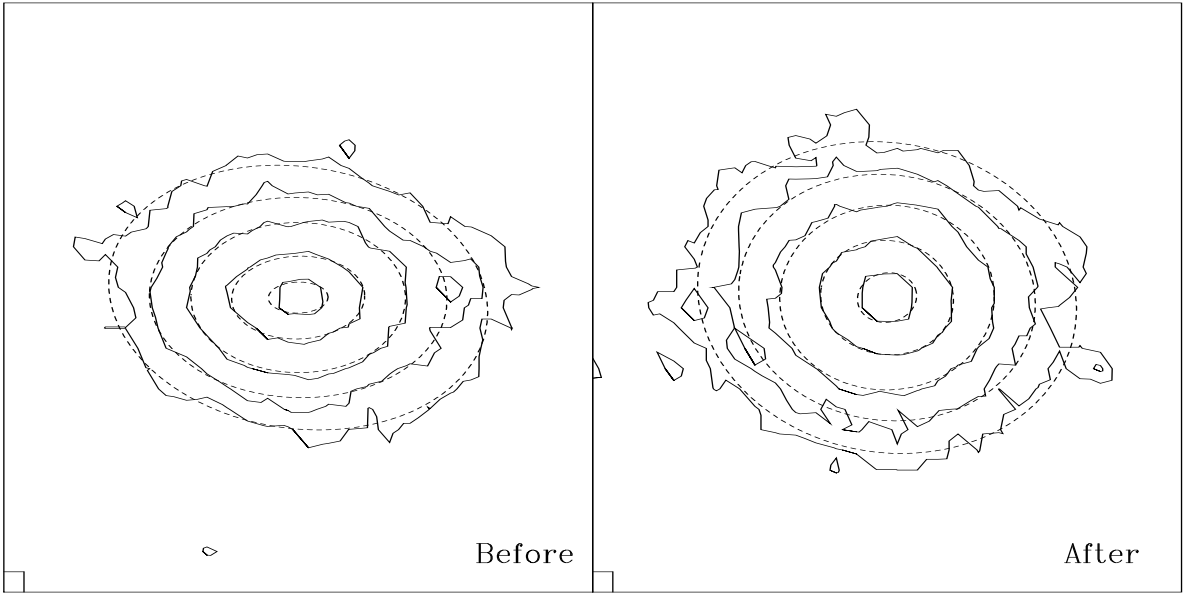

Fig. 5.—The surface density contours ("isophotes") of the dark halo in the plane of intermediate and major axis before and after the growth of the spherical Hernquist potential. The halo becomes considerably rounder in the equatorial plane and therefore more oblate.

We can demonstrate this effect by integrating representative box orbits in a flattened potential for the dark halo with a growing Hernquist potential to represent the baryonic component. For simplicity, we only look at orbits confined to a plane. We use the logarithmic potential for the flattened component given by,

$$\Phi = \frac{v_c^2}{2} \ln \left( r_L^2 + x^2 + \frac{y^2}{q_\Phi^2} \right) \tag{3}$$

where $q_\Phi$ is the flattening of the potential. The flattening of the density is, $\epsilon_\rho \approx 3\epsilon_\Phi$ where the ellipticity is $\epsilon = 1 - q$. We examine two models with $q_\Phi = 0.8$ and $q_\Phi = 0.9$ corresponding to $q_\rho \approx 0.4$ and $q_\rho \approx 0.7$. We scale the Hernquist potential so that $r_L/r_H = 3.0$ and $r_L v_c^2 / GM_H = 1.0$ corresponding approximately to the relative scaling of the N-body models.

The Hernquist model grows for 10 orbital periods for orbits with $r = r_L$. The elongation of the sample box orbits is chosen to match the axial ratio of the density profile. Orbits of this type will make up most of the mass distribution though some tubes are needed to fill in the space on the y axis.

Figure 6 presents the evolution of two sample box orbits in the logarithmic potentials with and without the growing sphere. The effect of the central component is striking. In both models, the elongation or axial ratio of the box orbits decreases. In the case of $q_\Phi = 0.9$, the orbit almost transforms into a loop orbit. There are very likely adiabatic invariants similar to the total angular momentum and radial action in spherical models that remain unchanged during the growth of the central potential (e.g. Blumenthal et al. 1986; Binney & May 1985). However, it is difficult to calculate them analytically in a general flattened potential. The orbits in the self-consistent N-body simulation are probably conserving similar quantities as they respond to the presence of a rounder potential. A full analysis of the transformation of orbital types is beyond the scope of this paper, though it would be interesting to examine the transformation of orbits selected from a flattened distribution function (e.g. Schwarzschild 1979; Statler 1987) in response to the growth of central galactic potential. The $q_\Phi = 0.9$ orbit almost changes into a loop orbit in response to the central potential while the change in the elongation of the $q_\Phi = 0.8$ orbit is not as great. One therefore expects that for an initially triaxial halo the ratio of $b/a$ will increase more than the ratio of $c/a$, therefore driving the halo to a more oblate form. The tendency to become oblate will be even stronger for disk potentials since the change in axial ratio will preferentially occur in the equatorial plane.

In summary, the growth of a central potential can change the shape of halo significantly driving an initially prolate/triaxial object to a rounder, more oblate object. The main effect probably results from the reduction of the elongation of box orbits as they readjust to the presence of the potential. In the following section, we examine the implications of rounder halos to galaxies.

## 4  Implications

### 4.1  Disk Ellipticity

A rotationally supported disk will be ovally distorted or elliptical if the potential is flattened in the disk plane (Kuijken & Tremaine 1991; Franx & de Zeeuw 1992). For a massless disk, the disk ellipticity is just the value for a closed loop orbit in the potential (assuming gaseous orbits do not cross). For a logarithmic potential, the ellipticity of a loop orbit is approximately equal to the ellipticity of the potential, i.e. $\epsilon_{loop} = \epsilon_\Phi$ although the long axis of the orbits are perpendicular to long axis of the potential. Because of this anti-alignment, if the disk mass is comparable to the halo mass within the disk the combined potential is considerably rounder. Disk ellipticities are therefore at most $\epsilon_\Phi$ but probably smaller. The ellipticity in the density is approximately related to ellipticity in the potential through, $\epsilon_\rho \approx 3\epsilon_\Phi$. If we assume that disks mainly lie in the $a - b$ plane of the halo so that there spin axes are aligned with the halo minor axis, we expect disk ellipticities at most equal to $\epsilon_{disk} = \epsilon_\Phi = \epsilon_\rho/3 \approx 0.10$ if dark halos have flattenings no smaller than $q_\rho = 0.7$ in the disk plane. When the disk mass is comparable to the halo mass, the disk ellipticity is even smaller, the resulting shape depending on the ratio of masses. Using the epicyclic approximation, for the combined model of a logarithmic potential and an axisymmetric exponential disk, the ellipticity of a loop orbit is

$$\epsilon_{loop} = \frac{2v_h^2}{2v_h^2 + v_d^2}\epsilon_\Phi \tag{4}$$

where $v_h$ and $v_d$ are the rotational velocities for the halo and disk components. If $v_d \approx v_h$ which we expect, then $\epsilon_{loop} = 2/3\epsilon_\Phi$. Of course, the disk is not axisymmetric but slightly elliptical so the net effect should be to reduce the disk ellipticity even further.

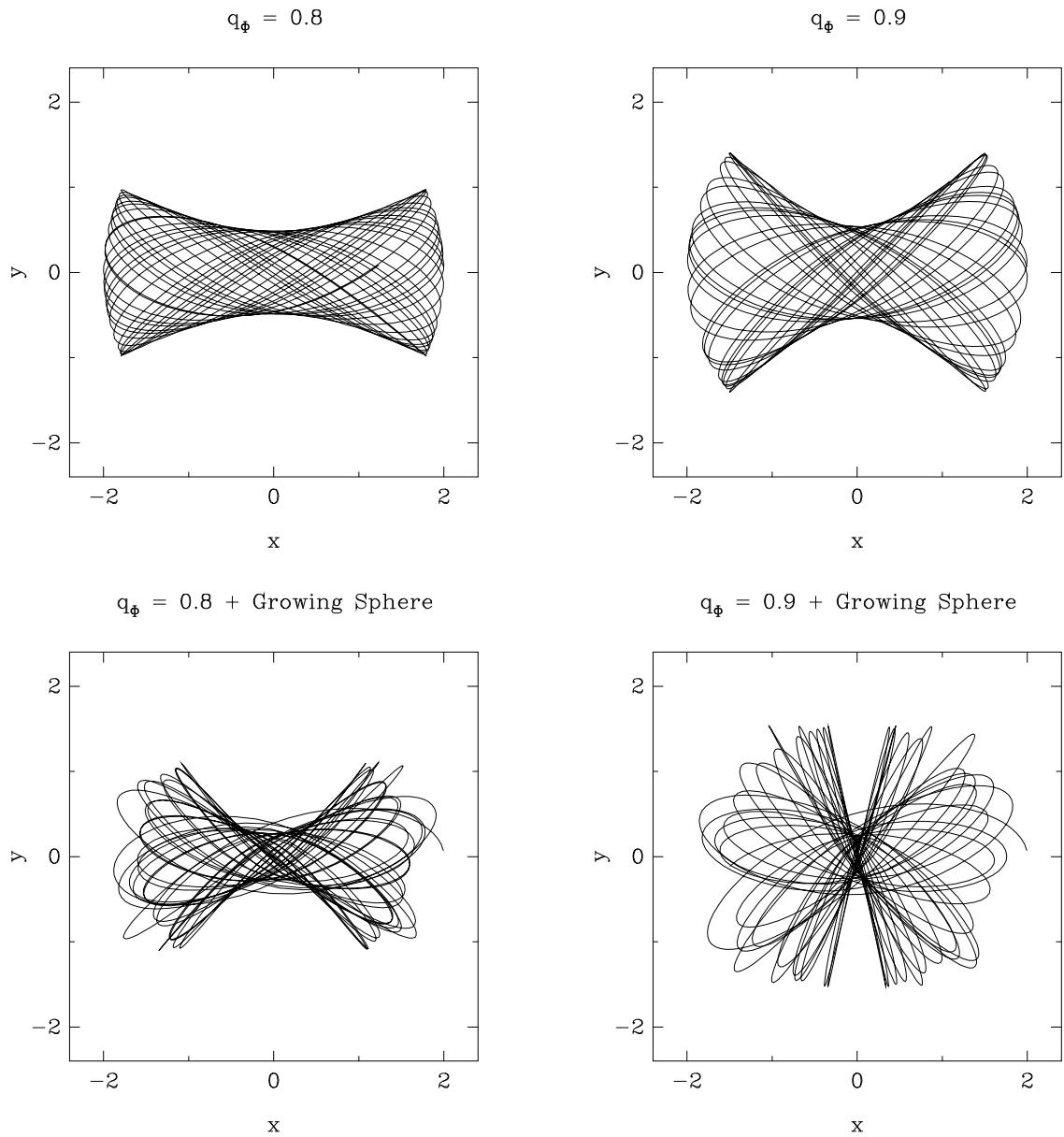

FIG. 6.–Sample box orbits in two flattened logarithmic potentials with $q_\Phi = 0.8$ and $0.9$ and the combined potentials including a growing central Hernquist sphere. Box orbits become less eccentric in response to the growing central potential.

There is evidence supporting a disk ellipticity around $\epsilon_{disk} = 0.10$ for spiral galaxies. Franx & de Zeeuw (1992) set an upper limit of $\epsilon_\Phi \approx 0.10$ by making the assumption that the scatter in the Tully-Fisher relation results solely from the assumption that disks are axisymmetric (while they are actually ovally distorted). Under the reasonable assumption that half the scatter comes from random errors, the upper limit is reduced to $\epsilon_\Phi = 0.06$. The distribution of apparent axial ratios of disk galaxies is fitted well by assuming that $\epsilon_{disk} = 0.10 - 0.13$ (Franx & de Zeeuw 1992). Kuijken & Tremaine (1993) also argue that the Galactic disk may have an ellipticity, $\epsilon_{disk} \approx 0.10$ to explain discrepancies in different methods of determining the velocity of the local standard of rest.

The results of the simulations suggest that $q_\rho \gtrsim 0.7 - 0.8$ in the plane of disk galaxies. If this is true then we expect disk ellipticities around $\epsilon_{disk} \lesssim 0.1$ though if disks contribute an approximately equal mass to the potential then $\epsilon_{disk} \lesssim 0.07$. The observed value around $\epsilon_{disk} = 0.10$ is consistent with the halo flattenings. If disks contribute significantly to the net potential and $\epsilon_{disk} = 0.1$ much flatter halos are still allowed with $q_\rho = 0.5$. Unfortunately, the disk ellipticity only poorly constrains the flattening of dark halo in the disk plane since it is weakly related to the halo flattening. However, the measured ellipticities are consistent with theoretical expectations.

## 4.2 The Intrinsic Shapes of Elliptical Galaxies

The shapes of dark halos arising in hierarchical collapse simulations with collisionless particles (Dubinski & Carlberg 1991; Warren et al. 1992) are intrinsically much flatter than the elliptical galaxies. The simulations in this paper suggest that dark halos may be much rounder than purely collisionless simulations might suggest if a dense lump of baryons gathers in the center of dark halo. Will elliptical galaxies also be rounder? The answer is not obvious since dissipation is the usual mechanism to describe the formation of disk galaxies. In hierarchical models, the formation of galaxies is not the simple two stage model of the formation of dark halo by hierarchical merging followed by the dissipation of the baryons. However, both these stages happen simultaneously. Dissipation occurs in every piece of substructure (perhaps at different times and at different rates) in the hierarchical collapse before the final merging and formation of the galaxy. One might imagine that disk galaxies result when the baryons in the substructure of the collapse remain gaseous until the time of merging, while elliptical galaxies form when some or most of the baryons in the substructure turn into stars before the time of merging. Nevertheless, during the chaotic phase of merging substructure the baryons sink to the center whether gaseous or stellar since the dense cores of galaxies in a merger make up the core of the merger remnant. We therefore expect a core of segregated baryons made of stars for elliptical galaxies. The stellar component of an elliptical galaxy in the early stages of its formation will not necessarily be distributed in the same shape or orientation as the dark halo because of the indirect segregation from dissipation in the substructure. We might therefore expect to see a similar effect on the shape of dark halos that we observed by growing disk and spherical potentials. Despite the many unknowns in the formation history of elliptical galaxies, we make the assumption (perhaps invalid) that the shapes of elliptical galaxies reflect the modified shapes of the dark halos under the influence of dissipation and explore the implications for the observed ellipticities and kinematics of elliptical galaxies.

This paper sets the new constraint that $b/a \gtrsim 0.8$ allowing us to construct a plausible function for the distribution of intrinsic shapes. The distribution should have the following properties:

1. The minimum value of $q_2 \equiv b/a \approx 0.8$. Since there is only a narrow range allowed for $q_2$ it is reasonable to assume that the distribution is uniform in the range $0.8 < q_2 < 1.0$.

2. The minimum flattening is $q_1 \equiv c/a = 0.3$ from stability arguments (Merritt & Stiavelli 1990; Merritt & Hernquist 1991). Also we observe no elliptical galaxies flatter than E7.

3. The distribution of flattenings is probably unimodal similar to a Gaussian distribution (Warren et al. 1992). Also the flattening is roughly preserved during baryonic infall.

With these various constraints, we can construct a hypothetical distribution function of shapes and then use it to fit the distribution of apparent ellipticities and kinematics for elliptical galaxies. This derived intrinsic distribution is generally not unique since by projecting the elliptical galaxies we effectively lose one of the dimensions of the distribution (see Statler 1993). Nevertheless, the distribution is physically motivated and will give us a useful consistency check between the distribution of elliptical galaxies and dark halos. Defining $\epsilon_1 = 1 - q_1$ and $\epsilon_2 = 1 - q_2$ as in Franx et al. (1991), a simple form of the distribution of intrinsic shapes is

$$P(\epsilon_1, \epsilon_2) = C \exp \left\{ -\frac{(\epsilon_1 - \mu_{\epsilon_1})^2}{2\sigma_{\epsilon_1}^2} \right\} \qquad (5)$$

with the constraints that $\epsilon_1 < \epsilon_2$, $\epsilon_2 < \epsilon_{2,max}$, and $\epsilon_1 < \epsilon_{1,max}$ and $C$ is a normalization constant. The distribution is uniform in $\epsilon_2$ and thus there is no direct dependence on $\epsilon_2$ in the function. The free parameters are $\mu_{\epsilon_1}$ and $\sigma_{\epsilon_1}$, the mean and dispersion in the halo flattening. The cutoff values of the distribution are $\epsilon_{1,max} = 0.7$ as determined by the observation of no elliptical galaxies flatter than E7 and $\epsilon_{2,max} = 0.2$ because of baryonic infall.

We now have a plausible distribution function of intrinsic shapes that we can use to fit to the observed distribution of ellipticities. We assume for simplicity that elliptical galaxies are perfect ellipsoids even though the observation of gradients in apparent ellipticity might suggest otherwise (e.g. Franx, Illingworth & Heckman 1989). After finding the best fit values of $\mu_{\epsilon_1}$ and $\sigma_{\epsilon_1}$ using maximum likelihood analysis, we can then test if these parameters are consistent with the measured distribution of halo flattenings from N-body simulations (Warren et al. 1992). We use two samples from the literature for the analysis: the distribution of 217 ellipticities from Franx et al. (1991) and Ryden's (1992) sample that is based on Djorgovski's (1985) sample. We divide these samples into 10 bins of width $\Delta\epsilon = 0.05$ except for the last one which contains all galaxies with $\epsilon > 0.45$. Following Ryden (1992), we find the most likely values for $\mu_{\epsilon_1}$ and $\sigma_{\epsilon_1}$ by minimizing,

$$\chi^2 = \sum_{i=1}^{N_b} \frac{(N_i - n_i)^2}{n_i} \qquad (6)$$

where $N_i$ is the observed number of galaxies in a bin and $n_i$ is the expected number given the assumed intrinsic shape distribution function. The distribution of apparent ellipticities, $g(\epsilon)$, is

$$g(\epsilon) = \iint_D P(\epsilon_1, \epsilon_2; \mu_{\epsilon_1}, \sigma_{\epsilon_1}) f(\epsilon \mid \epsilon_1, \epsilon_2) d\epsilon_1 d\epsilon_2 \qquad (7)$$

$f(\epsilon \mid \epsilon_1, \epsilon_2)$ is the conditional probability of apparent ellipticity given $\epsilon_1$ and $\epsilon_2$ (see Binney 1985 or Franx et al. 1991 for the derivation) and $D$ is the domain defined by $\epsilon_1 < \epsilon_{1,max}$, $\epsilon_2 < \epsilon_{2,max}$ and $\epsilon_2 < \epsilon_1$.

The most likely values from the analysis are $\mu_{\epsilon_1} = 0.32$ and $\sigma_{\epsilon_1} = 0.094$ for the Ryden sample (chi-squared probability $= 0.63$ with 8 degrees of freedom) and $\mu_{\epsilon_1} = 0.36$ and $\sigma_{\epsilon_1} = 0.13$ for the Franx et al. sample (chi-squared probability $= 0.42$). The modal value of $\mu_{\epsilon_1} \approx 0.35$ is the approximate value found from Franx et al.'s deprojection using Lucy's method (1974) when we assume that E galaxies are oblate. This is not surprising since we've constrained the shapes to have small $\epsilon_2$ or nearly oblate shapes. Figure 7 shows the 68% and 95% ($1\sigma$ and $2\sigma$) confidence intervals for both distribution calculated assuming $\chi^2$ follows a chi-squared distribution of degree 8. The broad $2\sigma$ confidence limits on $\sigma_{\epsilon_1}$ suggest that a model in which the ellipticity distribution is uniform over the given domain is as plausible as the proposed Gaussian distribution. Figure 8 illustrates the good quality of the fit to the observations for the Franx sample. Franx's sample predicts an intrinsically flatter distribution than the Ryden sample, though the $1\sigma$ confidence intervals overlap for the distributions. Warren et al. 1991 have the largest sample of dark halos for comparison to the galaxy flattenings. Their distribution of halo flattenings peaks around $\epsilon_1 = 0.55$ at 10 kpc and $\epsilon_1 = 0.45$ at 40 kpc with a dispersion $\approx 0.15$ (see their Figure 7b). The dispersion in flattening is similar though the halos are still slightly flatter than the inferred distribution of elliptical galaxy shapes. We found that a central spherical potential also managed to increase the flattening slightly so this may be adequate to make up the difference of 0.10 mean value of $\mu_{\epsilon_1}$.

### 4.3 Minor Axis Rotation in Elliptical Galaxies

We can also check if this hypothetical shape distribution is consistent with the observations of minor axis rotation in elliptical galaxies. We assume that the rotation of the elliptical galaxies is about the intrinsic minor axis as in the dark halos (Warren et al. 1992; Dubinski 1992). With this assumption, we can construct the distribution function of kinematic misalignment angle, $\Psi$, as defined by Franx et al. (1991). The angle $\Psi$ is just the angle between the apparent minor axis and the projected axis of rotation. Observationally, the angle is approximated as $\tan \Psi = v_{min}/v_{maj}$ where $v_{min}$ and $v_{maj}$ are the maximum rotational velocities measured along the apparent minor and major axes of the observed galaxy. Figure 9 presents the distribution in $\Psi$ assuming the galaxies rotate intrinsically about their minor axes that can be compared to Franx et al.'s Figure 17a. The distribution is sharply peaked at $\Psi = 0°$ levelling off to a constant value for $\Psi \gtrsim 30°$. A total of 17% of galaxies are misaligned by more than 30° compared with 26% for Franx's et al.'s sample (9 of 35 galaxies with angle errors less than 25°). This is in reasonable agreement though there is a slight peak of 4 galaxies at $\Psi = 90°$ bin which is not consistent with the inferred distribution. The number of galaxies is small in this bin however, so it may be reflecting statistical error. If the peak is real, the implication is that there is probably some fraction of the population that rotate about their intrinsic major axes (e.g. Ryden 1992).

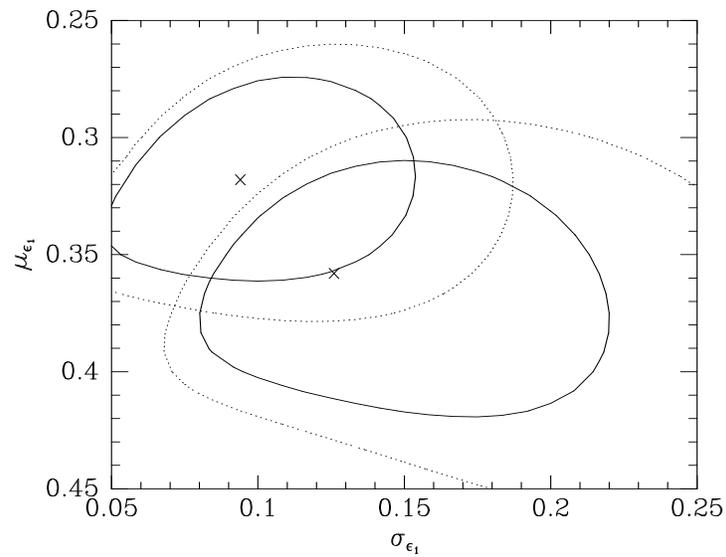

FIG. 7.–The most likely values for $\mu_{\epsilon_1}$ and $\sigma_{\epsilon_1}$ along with the 1 and $2\sigma$ confidence intervals as determined by a $\chi^2$ test. The contours to the upper left are derived from Ryden's (1992) sample while the other contours are derived from Franx et al.'s (1991) sample.

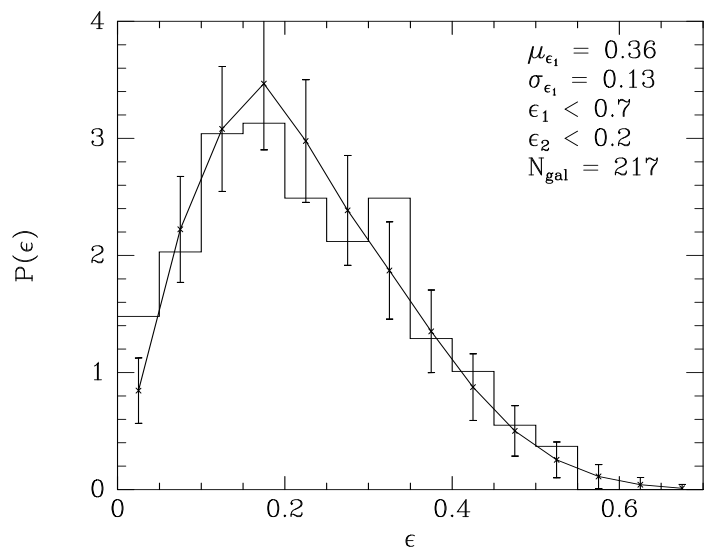

FIG. 8.–The best fit distribution to the distribution of apparent ellipticities for the Franx et al. (1991) sample.

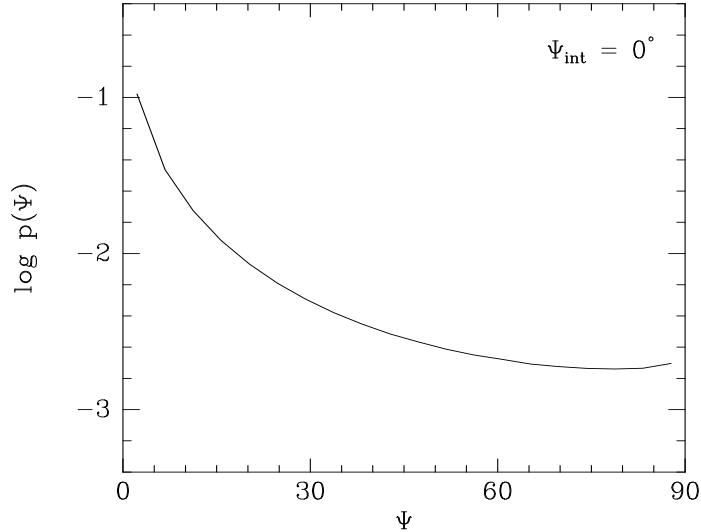

FIG. 9.–The distribution of kinematic misalignment angle, $\Psi$ for the intrinsic distribution of shapes with $\mu_{\epsilon_1} = 0.64$ and $\sigma_{\epsilon_1} = 0.13$ (Franx et al. (1991) sample). A total of 17% of galaxies are misaligned by more than 30° despite an intrinsic alignment about the minor axis.

## 5 CONCLUSIONS

The formation of the luminous galaxy by dissipative infall can modify the orbits of dark halos changing the density profile and shape. When a disk forms in the center with its spin axis aligned with the halo minor axis, the axial ratio $b/a$ can grow from $b/a = 0.5$ to $b/a = 0.7 - 0.8$. The flattening $c/a$ does not change significantly. Dark halos around spiral galaxies are probably more oblate than the predictions of dissipationless simulations. When a spherical potential grows in the center, the halo generally becomes rounder as well with the largest growth in the axial ratio happening in the plane of the middle and major axis ($a - b$ plane). Again, halos become more oblate while approximately preserving their flattening although the $c/a$ increases slightly as well.

The net effect is that disk ellipticities should not be as pronounced as originally predicted though values as large as $\epsilon_{disk} = 0.07$ are permitted. If we assume that elliptical galaxies have the same modified shape distribution as dark halos (i.e. $b/a \gtrsim 0.8$ and $c/a$ unchanged) the inferred distribution of apparent ellipticities is in much better agreement with the observed distribution. Halos are still slightly more flattened however. The distribution of misalignment angles also is in fair agreement with the observed distribution when we assume that galaxies rotate about their intrinsic minor axis. The observed peak of 4 galaxies at $\Psi \approx 90°$ in the Franx et al. sample may imply a subpopulation of galaxies that rotate about their intrinsic major axis.

## Acknowledgments

I would like to thank Ray Carlberg, Konrad Kuijken and Marijn Franx for useful discussions and comments. I also thank Avi Loeb, Chris Kochanek and Eyal Maoz for additional comments. I acknowledge the financial support of a Center for Astrophysics Postdoctoral Fellowship.

## REFERENCES

Bardeen, J.M., Bond, J. R., Kaiser, N., & Szalay, A. S. 1986, ApJ, 304, 15 [BBKS]

Barnes, J., & Hut, P. 1986, Nature, 324, 446

Barnes & White 1984, MNRAS, 211, 753

Bertola, F., & Capaccioli, M. 1975, ApJ, 200, 439

Bertschinger, E. & Jain, B. 1993, preprint

Binney, J. J. 1978, MNRAS, 183, 501

Binney, J. 1985, MNRAS, 212, 767

Binney, J. J., & May, A. 1986, MNRAS, 218, 743

Binney, J., & Tremaine, S. 1987, Galactic Dynamics, (Princeton: Princeton University Press)

Blumenthal, G. R., Faber, S. M., Flores, R., & Primack, J. R. 1986, ApJ, 301, 27.

Djorgovski, S. 1986, Ph.D. thesis, Univ. of California, Berkeley

Dubinski, J. 1992, ApJ, 401, 441

Dubinski, J.& Carlberg, R. 1991, ApJ, 378, 496

Evrard, A. E., Summers, F. J., & Davis, M. 1993, preprint

Flores, R., Primack, J. R., Blumenthal, G. R., Faber, S. M., 1993, ApJ, 412, 443

Franx, M. & de Zeeuw, T. 1992, ApJL, L47

Franx, M., Illingworth, G., & de Zeeuw, T. 1991, ApJ, 383, 112

Franx, M., Illingworth, G., & Heckman, T. 1989, AJ, 98, 538

Frenk, C. S., White, S. D. M., Davis, M., & Efstathiou, G., 1988, ApJ, 327, 507

Hernquist, L. 1990, ApJ, 356, 359

Hernquist, L. & Katz, N. 1989, ApJS, 70, 419

Jedrzejewski, R., & Schechter, P. 1989, AJ, 98, 147

Katz, N. & Gunn, J. E. 1991, ApJ, 377, 365

Katz, N., Quinn, T., & Gelb, J. 1992, preprint

Kuijken, K., and Gilmore, G. 1989, MNRAS, 239, 651

Kuijken, K. & Tremaine, S. 1993, preprint

Kuijken, K. & Tremaine, S. 1992, in Dynamics of Disk Galaxies, e. B. Sundelius (Goteborg: Goteborg Univ. Press), 71

Kuijken, K. 1991, ApJ, 376, 467

Rees, M. J. & Ostriker, J. P. 1977, MNRAS, 179, 541

Lucy, L. B. 1974, AJ, 79, 745

Merritt, D., & Stiavelli, M. 1990, ApJ, 358, 399

Merritt, D., & Hernquist, L. 1991, ApJ, 376, 439

Miyamoto, M., & Nagai, R. 1975, PASJ, 27, 533

Quinn, P. J., Salmon, J. K., & Zurek, W.H. 1986, Nature, 322, 329

Ryden, B. S., & Gunn, J.E. 1987, ApJ, 318, 15

Ryden, B. S., 1988, ApJ, 329, 589

Ryden, B. S. 1991, ApJ, 370, 15

Ryden, B. S. 1992 , ApJ, 396, 445

Sackett, P. D., & Sparke, L. S. 1990, ApJ, 361, 408

Saglia, R. P. et al. 1993, ApJ, 403, 567

Schwarzschild, M. 1979, ApJ, 232, 236

Statler, T. S. 1987, ApJ, 321, 113

Statler, T. S. 1993, preprint

Sanchez-Saavedra, M. L., Battener, E. and Florido, E. 1990, MNRAS, 246, 458

Warren, M. S., Quinn, P. J., Salmon, J. K., & Zurek, W. H. 1992, ApJ, 399, 405

White, S. D. M., & Rees, M. J. 1978, MNRAS, 183, 341

White, S. D. M., & Frenk, C. S. 1991, ApJ, 379, 52

Whitmore, B. C., McElroy, D. B., & Schweizer, F. 1987, ApJ, 314, 439

Zel'dovich, Ya. B. 1970, AA, 5, 84